\documentstyle[psfig]{mn}

\newif\ifAMStwofonts

\def\simgt{\hbox{\rlap{\raise 0.425ex\hbox{$>$}}\lower 0.65ex\hbox{$\sim$}}}
\def\simlt{\hbox{\rlap{\raise 0.425ex\hbox{$<$}}\lower 0.65ex\hbox{$\sim$}}}
\def\h1{$h^{-1}$}

\ifoldfss
  \ifCUPmtlplainloaded \else
    \NewTextAlphabet{textbfit} {cmbxti10} {}
    \NewTextAlphabet{textbfss} {cmssbx10} {}
    \NewMathAlphabet{mathbfit} {cmbxti10} {} 
    \NewMathAlphabet{mathbfss} {cmssbx10} {} 
  \fi
  \ifAMStwofonts
    \ifCUPmtlplainloaded \else
      \NewSymbolFont{upmath} {eurm10}
      \NewSymbolFont{AMSa} {msam10}
      \NewMathSymbol{\upi}     {0}{upmath}{19}
      \NewMathSymbol{\umu}     {0}{upmath}{16}
      \NewMathSymbol{\upartial}{0}{upmath}{40}
      \NewMathSymbol{\leqslant}{3}{AMSa}{36}
      \NewMathSymbol{\geqslant}{3}{AMSa}{3E}

    \fi
  \fi
\fi 

\ifnfssone
  \newmathalphabet{\mathit}
  \addtoversion{normal}{\mathit}{cmr}{m}{it}
  \addtoversion{bold}{\mathit}{cmr}{bx}{it}
  \newmathalphabet{\mathbfit} 
  \addtoversion{normal}{\mathbfit}{cmr}{bx}{it}
  \addtoversion{bold}{\mathbfit}{cmr}{bx}{it}
  \newmathalphabet{\mathbfss} 
  \addtoversion{normal}{\mathbfss}{cmss}{bx}{n}
  \addtoversion{bold}{\mathbfss}{cmss}{bx}{n}
  \ifAMStwofonts
    \ifCUPmtlplainloaded \else
      \UseAMStwoboldmath
      \makeatletter
      \new@mathgroup\upmath@group
      \define@mathgroup\mv@normal\upmath@group{eur}{m}{n}
      \define@mathgroup\mv@bold\upmath@group{eur}{b}{n}
      \edef\UPM{\hexnumber\upmath@group}
      \new@mathgroup\amsa@group
      \define@mathgroup\mv@normal\amsa@group{msa}{m}{n}
      \define@mathgroup\mv@bold\amsa@group{msa}{m}{n}
      \edef\AMSa{\hexnumber\amsa@group}
      \makeatother
      \mathchardef\upi="0\UPM19
      \mathchardef\umu="0\UPM16
      \mathchardef\upartial="0\UPM40
      \mathchardef\leqslant="3\AMSa36
      \mathchardef\geqslant="3\AMSa3E
    \fi
  \fi
\fi 
\ifnfsstwo
  \DeclareMathAlphabet{\mathbfit}{OT1}{cmr}{bx}{it}
  \SetMathAlphabet\mathbfit{bold}{OT1}{cmr}{bx}{it}
  \DeclareMathAlphabet{\mathbfss}{OT1}{cmss}{bx}{n}
  \SetMathAlphabet\mathbfss{bold}{OT1}{cmss}{bx}{n}
  \ifAMStwofonts
    \ifCUPmtlplainloaded \else
      \DeclareSymbolFont{UPM}{U}{eur}{m}{n}
      \SetSymbolFont{UPM}{bold}{U}{eur}{b}{n}
      \DeclareSymbolFont{AMSa}{U}{msa}{m}{n}
      \DeclareMathSymbol{\upi}{0}{UPM}{"19}
      \DeclareMathSymbol{\umu}{0}{UPM}{"16}
      \DeclareMathSymbol{\upartial}{0}{UPM}{"40}
      \DeclareMathSymbol{\leqslant}{3}{AMSa}{"36}
      \DeclareMathSymbol{\geqslant}{3}{AMSa}{"3E}
    \fi
  \fi
\fi 

\ifCUPmtlplainloaded \else
  \ifAMStwofonts \else 
    \def\upi{\pi}
    \def\umu{\mu}
    \def\upartial{\partial}
  \fi
\fi

\title[Microlensing of BAL Quasars]{Microlensing of Broad Absorption Line
Quasars} \author[G. F. Lewis and
K. E. Belle] { G. F. Lewis\thanks{
Present Address: Astronomy Dept., University of
Washington, Box 351580, Seattle, WA 98195-1580, U.S.A.  \& Dept.
of Physics and Astronomy, University of Victoria, PO Box 3055,
Victoria, BC V8W 3P6, Canada }\thanks{
Email: {\bf \tt
gfl@astro.washington.edu}} and K. E. Belle\thanks{Email: {\bf \tt
kbelle@mail.ess.sunysb.edu}} \\ Astronomy Group, Dept. of Earth and
Space Sciences, \\ SUNY at Stony Brook, NY 11794-2100. U.S.A }
\date{Zeroth Draft} \pubyear{1997}

\begin{document}

\maketitle

\newcommand{\fmmm}[1]{\mbox{$#1$}}
\newcommand{\scnd}{\mbox{\fmmm{''}\hskip-0.3em .}}
\newcommand{\scnp}{\mbox{\fmmm{''}}}

\begin{abstract}
The physical nature of the material responsible for the
high--velocity, broad--absorption line features seen in a small
fraction of quasar spectra has been the subject of debate since their
discovery.  This has been especially compounded by the lack of
observational probes of the absorbing region. In this paper we examine
the r\^{o}le of ``microlenses'' in external galaxies on observed
variability in the profiles of broad absorption lines in
multiply-imaged quasars. Utilizing realistic models for both the broad
absorption line region and the action of an ensemble of microlensing
masses, we demonstrate that stars at cosmological distances can
provide an important probe of the physical state and structure of
material at the heart of these complex systems.  Applying these
results to the macrolensed BAL quasar system, H1413+117, the observed
spectral variations are readily reproduced, but without the
fine--tuning requirements of earlier studies which employ more
simplistic models.
\end{abstract}

\begin{keywords}
Gravitational Lensing, Microlensing, BAL: QSOs
\end{keywords}

\section{Introduction}

Microlensing in extragalactic systems has been seen not only to
introduce rapid fluctuations into the light curves of macrolensed
quasars~\cite{irwin1989,corrigan1991}, but also to introduce
significant variability into their spectral
features~\cite{filippenko1989,lewis1997}.  Through the spectroscopic
monitoring of such variations in microlensed quasars, the kinematics
and scales of structure in their inner regions can be
probed~\cite{kayser1986,nemiroff1988,schneider1990}.

In this paper, the question of whether a peculiar class of quasars,
those exhibiting Broad Absorption Line (BAL) features, can be
microlensed in a similar fashion is addressed, and whether this
microlensing can shed light on the physics of the material at the core
of these quasars.  A description of the background of the problem is
presented, with reference to previous study.  An outline of the
simple, but realistic, model employed in this study is given and 
the action of microlensing at a fold caustic is described.  The
results of this study demonstrate that microlensing can provide a
valuable probe of BAL quasars.

\section{Background}\label{background}

\subsection{Microlensing}\label{micro}

Soon after the discovery of the first gravitationally lensed
system~\cite{walsh1979}, Chang and Refsdal (1979) noted that the
small--scale granularity in galactic matter distributions, the fact
that galaxies are comprised of point--like stars rather than solely
smoothly distributed material, would lead to a peculiar gravitational
lensing effect. Although the image splitting by such stellar mass
objects, $\left(\sim10^{-6}\scnp\right)$, would be too small to be
observable, the induced amplification by these ``microlenses'' could
be substantial. As the degree of amplification is dependent upon the
relative position of the source and microlensing star, individual
stellar motions can introduce a significant time--dependent variability
into the light curve of a distant source.

The original work of Chang and Refsdal~(1979) considered the action of
a single microlensing star on the light curve of a macrolensed quasar.
The development of numerical techniques demonstrated that, with an
ensemble of stars, the microlensing action combines in a highly
non--linear fashion to introduce complex structure into a microlensed
light curve~\cite{young1981,paczynski1986a}. The implementation of the
backwards ray--shooting technique~\cite{kayser1986,wambsganss_thesis},
and more recently the contour algorithm~\cite{witt1993,lewis1993},
allowed an analysis of the statistical properties of the microlensing
induced variability (Lewis \& Irwin 1995; 1996).

At the source, the microlensing scale length, the Einstein radius, for
a solar mass star is given by~\cite{schneiderbook},
\begin{equation}
\eta_o = \sqrt{ \frac{\rm 4 G M_{\odot}}{\rm c^2} \frac{D_{ls} D_{os}}
{D_{ol}}}\ ,
\label{er}
\end{equation}
Here, $D_{ij}$ are the angular diameter distances between an observer
(o), lens (l) and source (s), respectively.  For typical multiply
imaged quasars this characteristic scale is $\sim 0.01-0.1$pc.

Numerical simulations demonstrated that, at even low optical depths,
the induced variability of a microlensed source can be extremely
complex. This is due to the microlensing amplification possessing a
strong dependence on the relative positions of the source quasar and
microlensing stars which results in a similarly complex map of
amplification over the source plane~\cite{wambsganss_thesis}.  This
map possesses an intricate web of lines of formally infinite
amplification which are known as caustics.  As these cross an extended
source, dramatic flux variations can be observed~\cite{chang1984}. In
the absence of a stellar velocity
dispersion~\cite{schramm1993,wambsganss1995}, the characteristic
caustic crossing--time of a source is
\begin{equation}
\tau \sim \frac{ f_{\rm 15} }{ 1 + z_l } \frac{ D_{ol} }{ D_{os} }
\frac{ h_{75} } { v_{300} } yrs,
\label{timescale}
\end{equation}
where the scale--size of the source is $f_{\rm 15}\times10^{15}~{\rm
h_{75}}$cm and $z_l$ is the redshift of the lensing
galaxy~\cite{kayser1986}.  The velocity of the microlensing stars
across the line--of--sight is 300$v_{300}$km~s$^{-1}$.  In observed
macrolensed systems, $\tau\sim$ several months.

\subsection{BAL Quasars$^{\P}$}
~\footnotetext[5]{Although technically ``BAL quasars'', as they tend
to be radio-quiet, should be identified as BAL QSOs, we refer to them
as quasars for clarity.}

BAL quasars, representing 3-10\% of the quasar population, are
identified as possessing spectra with resonance absorption line systems
due to highly ionized species. Although associated with the source
quasar, these lines exhibit bulk outflow velocities of 5000--30000
${\rm km\;s^{-1}}$~\cite{turnshek1984}.  

Several recent reviews have discussed the nature of these absorption
line systems~\cite{turnshek1988,turnshek1995,weymann1995}.  The
following is drawn from these articles.  Current theories present two
possible models for the rarity of BAL quasars: the BAL covering
fraction is close to unity such that all lines of sight would result
in BAL systems, with only 3-10\% of quasars possessing a BAL region,
or, BAL absorption systems are ubiquitous, although the covering
factor is small, equivalent to the percentage of quasars observed with
BALs, therefore making the detection of BALs dependent upon the line
of sight to the quasar core.  However, it has been proposed that the
size of the BAL covering factor could differ between individual
quasars, resulting in differences of the observed BAL profiles.

Although the geometry of the BAL region has not been clearly defined
through observations, polarization measurements and inferred electron
densities have placed constraints on the scale--size of the BAL
region. Namely, the region must lie within 30--500 pc from the central
continuum source over the broad range of quasar luminosities.  There
is also evidence to suggest that the large scale geometry represents a
disk, while locally, the individual clouds are confined to flat
disk-like shapes whose surfaces are perpendicular to the plane of the
entire disk~\cite{turnshek1986}.  Column density measurements yield a
radial extent of $\sim10^{11} - 10^{13.5}\rm{cm}$ for these
clouds~\cite{turnshek1995,murray1995}, although scale lengths as
small as $10^8$cm have been suggested~\cite{weymann1995}.

Until recently, all observed BAL quasars have been classified as weak
radio sources, suggesting that these and radio-loud quasars are drawn
from mutually exclusive populations.  However, Becker et al. (1997)
recently identified the first known radio-loud BAL quasar.  Lacking
the typically strong characteristic quasar emission lines, this object
possesses nearly saturated BALs, and a highly absorbed continuum,
suggesting that the radio luminosity is possibly related to the degree
of absorption to the object.

Observations have revealed that BAL quasars are more highly polarized
than radio quiet non-BAL quasars~\cite{moore1984}.  It has been
suggested that in BAL quasars, the continuum flux is attenuated upon
leaving the continuum source as it passes through a broad emission
line (BEL) region, a BAL region, or a scattering
region~\cite{goodrich1995}.  Higher polarized flux is due to
scattering farther from the BEL
clouds~\cite{goodrich1995,goodrich1997}.  A recent polarization
study~\cite{ogle1997} suggests that several different geometries of
the distribution of BAL material could account for the polarization
observed in several BAL quasar spectra, and that an equatorial outflow
within the BAL region is the most probable case.

Many models attempt to explain the origin and confinement of the BAL
region material: such as magnetically confined clouds~\cite{arav1994}
and x-ray--shielded flows~\cite{murray1995}.  The small distances from
the central continuum source, coupled with high outflow velocities,
imply short time--scales for the crossing of the individual clouds
through the entire BAL region, as compared with the quasar lifetime.
The constant source of the material which is necessary to maintain the
BAL region could originate from several sources; including atmospheres
of giant stars~\cite{scoville1995}, novae~\cite{shields1996}, parts of
an inner obscuring torus, or parts of the accretion
disk~\cite{murray1995}.

\subsection{H1413+117}

Serendipitously discovered in a survey of highly luminous quasars,
H1413+117 consists of four images of a $z=2.55$ quasar, arranged in a
cruciform with angular separations of $0\scnd77$ to
$1\scnd36$~\cite{magain1988}. Although no individual lensing candidate
has been identified in deep imaging~\cite{lawrence1996}, Magain et
al. (1988) noted that a number of absorption systems, at $z\sim
1.4-2.1$, could indicate the presence of several lensing
candidates. This is borne out with the recent study of Kneib et
al. (1997) who detected the presence of a ``cluster'' in front of
H1413+117 in HST WFPC2 images, which they associate with these
absorption systems.  With the addition of CO
observations~\cite{barvainis1994,barvainis1997,yun1997} and the
existence of the foreground cluster, Kneib et al. (1997) attempt to
improve previous lensing models~\cite{kayser1990}.  This additional
cluster potential, coupled with a modest mass lensing galaxy, is seen
to accurately reproduce the observed system.
  
Spectroscopic observations of H1413+117 were undertaken by Angonin et
al. (1990).  As well as confirming the BAL nature of the source, it
was noted that differences existed between the spectra of the
individual images.  Most noticeably, the relative strengths of the
Si~IV~$\lambda\lambda 1394,1403 $ and C~IV~$\lambda 1549$ emission
lines vary between the images~\cite{hutsemekers1993}, a signature of
differential amplification due to microlensing by stellar mass
objects~\cite{sanitt1971,lewis1997}.  Coupled with this, it was seen
that in one of the images the associated BAL profiles exhibit
significant differences of $\sim20\%$ when compared to the other
images.

Motivated by the BAL profile differences in the spectra of H1413+117,
Hutsem\'ekers (1993, Hutsem\'ekers et al. 1994) undertook an
investigation of the possible microlensing effects of stars on 
BAL profiles.  This work examines the effects on an individual cloud,
a sub-region with differing optical depth from the average of the
entire BAL region, due to a single microlens.

Considering an isolated point mass microlens~\cite{einstein1936},
Hutsem\'ekers (1993) shows that the maximum resultant amplification is
less than that of the observed spectral differences.  The
Chang-Refsdal lens (Chang \& Refsdal 1979; 1984) considers the
non-linear microlensing effects of a point mass located in a galactic
shear, whereby critical curves and caustics will dictate the
amplification effects.  This model is able to reproduce the
amplification seen within the spectra of the multiple images of
H1413+117~\cite{hutsemekers1993}.

It was concluded that the observed spectral variations in the BAL
profiles of H1413+117 could be a result of the selective
magnification due to microlensing, but only for a specific,
``fine-tuned'' configuration; the cloud must be at the border of the
BAL region and lie along a caustic, while the remainder of the BAL
region must lie in an area of strong de-magnification.  This model
relies strongly on the distance between the projected center of the
lens and the center of the BAL region; a slight change will greatly
effect the resulting magnification of the
region~\cite{hutsemekers1993}.

\section{A Model for the BAL Region}\label{model}

For the analysis presented here the quasar continuum source was
modeled as a 2--dimensional Gaussian surface--brightness of radius
$10^{15}~{\rm h_{75}^{-1}~cm}$~\cite{rees1984}. The material producing
the BAL absorption is placed along the line-of--sight to the continuum
source and the surface--brightness of this source as viewed through
this region, as a function of wavelength, was calculated.

\begin{figure}
\centerline{
\psfig{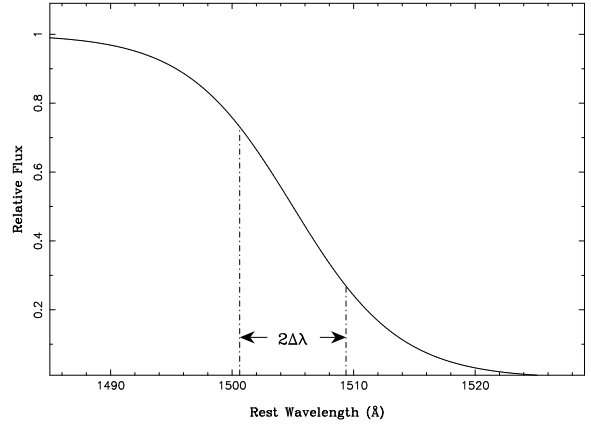}
}	
\label{sourcep}
\caption[]{ 
The adopted profile of the Broad Absorption Line, as given by
Equation (3). The parameters of this profile were chosen
to represent absorption by C~IV~$\lambda 1549$.  }
\end{figure}

\begin{figure*}
\centerline{
\psfig{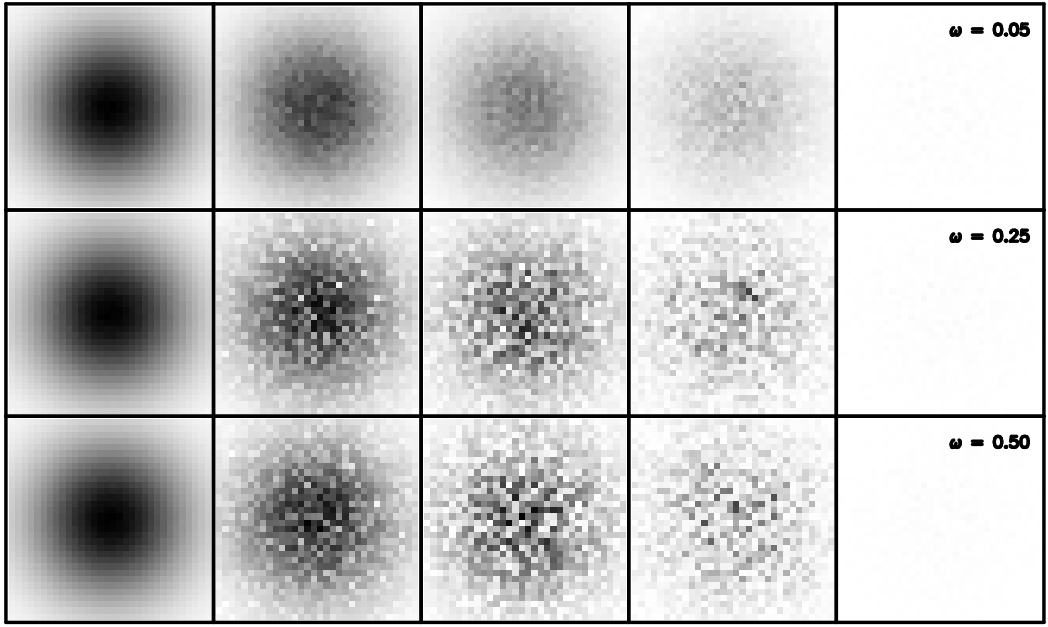}
}
\label{source}
\caption[]{ 
Running from left--to--right, these panels show the quasar continuum
source at $1\%$, $25\%$, $50\%$, $75\%$ and $99\%$ absorption,
respectively. From top--to--bottom, the panels represent absorption
widths of $\omega = 5\%$, $25\%$ \& $50\%$, respectively. In this
example, the absorption clouds are $1.2\times10^{14} {\rm h_{75}^{-1}cm}$ 
in extent, while the entire source region is  $3.7\times10^{15} 
{\rm h_{75}^{-1}cm}$.}
\end{figure*}

The absorption profile of the material within the BAL region was
modeled with a simple functional form, namely
\begin{equation}
{\rm A}\left(\lambda\right) = 
\frac{1}{1 + \exp{ \frac{ \lambda - \lambda_c }{\Delta\lambda} }}
\label{absprofile}
\end{equation}
where $\lambda_c$ is the wavelength at which $50\%$ of the light from
the continuum core is absorbed, while $2\Delta\lambda$ is the
``width'' of the absorption feature between $73\%$ and $27\%$
absorption. The absorption profile was chosen to represent the
blue--wing of C~IV~$\lambda1549$, with $\lambda_c = 1505{\rm \AA}$,
corresponding to a bulk outflow at $50\%$ absorption of $\sim8500~{\rm
km~s^{-1}}$. With an absorption width of $2~\Delta\lambda = 8.75{\rm
\AA}$, the resulting profile [Figure (1)] adequately
represents observed C~IV~$\lambda 1549$ BAL
profiles~\cite{turnshek1995,weymann1995}.

The over--all BAL region is characterised by two quantities; a
linear measure describing the scale--length of inhomogeneities in the
absorbing material (henceforth referred to as the ``cloud--size''), and the
degree of variation of absorption at a particular wavelength. The
distribution in absorption is represented by a Gaussian of width
$\omega$, between 0\% and 100\% absorption, centred at
$A_c\left(\lambda\right)$.  This value was chosen such that the mean
absorption, $\left< A\left(\lambda\right) \right>$, drawn from this
distribution is the value given by Equation~\ref{absprofile}. Note
that if, for a particular $\left< A\left(\lambda\right) \right>$ and
$\omega$, that $A_c\left(\lambda\right)$ lies outside the 0\%
$\rightarrow$ 100\% boundary, then $A_c\left(\lambda\right)$ is fixed
at the boundary position and $\omega$ is varied to ensure $\left<
A\left(\lambda\right) \right>$ follows Equation~\ref{absprofile}.

For this study we select the observational parameters of the quadruply
lensed quasar H1413+117~\cite{angonin1990}, with the source quasar
being at $z=2.55$. Although the redshift of the lensing galaxy has yet
to be established, there is evidence to suggest it is
$z\sim1.55$. This value is adopted here, resulting in an Einstein
radius in the source plane, for a solar mass star, (Equation~\ref{er}) of
$1.9\times10^{16}{\rm h_{75}^{-\frac{1}{2}}}$cm (Throughout we
assume $\Omega = 1$ and $\Lambda = 0$).

Figure (2) presents an example of a quasar source as observed through
the BAL region absorption clouds. The ``source region'', $r_{s}$, is
3.7 source radii in extent and the clouds are laid out on a
$32\times32$ grid with a resulting scale--size of
$1.2\times10^{14}~{\rm h_{75}^{-1}~cm}$.  From left--to--right the
panels present absorption of 1\%, 25\%, 50\%, 75\% and 99\%
absorption, while from top--to--bottom $\omega$ = 5\%, 25\% and 50\%
respectively.

For the analysis presented here, we chose cloud sizes of
$2^{-n}r_{s}, n=1..9$.  With this, the largest clouds were
$1.8\times10^{15}{\rm h_{75}^{-1}~cm}$ in extent while the smallest
were $7.1\times10^{12}{\rm h_{75}^{-1}~cm}$.

\section{Microlensing Model}\label{fold}

\subsection{Amplification at fold caustics}

At low to medium optical depths the scale of the BAL region is small
compared to the typical scales of caustics produced by
microlensing. Within this regime it can be assumed that the majority
of caustics crossing the BAL source will be isolated fold
catastrophes~\cite{schneiderbook}. Also, as the scale of the BAL
source is expected to be much less than the caustic curvature it can
be assumed that within the region of interest the caustic is
locally straight. With this, the amplification of a point source at $y$,
relative to a straight fold caustic at $y_c$ is given by~\footnote{A
full description of the action of gravitational lensing near caustic
critical points is presented in Chapter 6 of ``Gravitational Lenses''
by Schneider et al. (1992)}
\begin{equation}
\mu_p( y ) = \sqrt{ \frac{ g }{ y - y_c} } H( y - y_c ) + \mu_o .
\label{pointamp}
\end{equation} 
Here, $y - y_c$ is the perpendicular distance between the point of
interest and the caustic line, $H(y)$ is the Heaviside step function
and $\mu_o$ is the amplification due to all other images of the
source. This is assumed to be constant in the vicinity of the
caustic. The parameter $g$ represents the ``strength'' of the caustic.

The amplification, or magnification, of an extended source is given
by
\begin{equation}
\mu = \frac{ \int\!\!\int I({\bf y })\,\mu_p({\bf y})\,d^2y}
{ \int\!\!\int I({\bf y })\,d^2y} ,
\label{extend}
\end{equation}
where $I({\bf y })$ is the surface brightness distribution of the
source.  If the source is considered to be a square ``pixel'' of
uniform surface brightness which is aligned such that its sides are
parallel/perpendicular to the caustic line, and it has its nearest 
edge a distance $y_k$ from a caustic of strength $g$, it will be
magnified by a factor
\begin{equation}
\mu_{pix} = 2 \frac{\sqrt{g}}{\Delta y}
\left( \sqrt{ y_k + \Delta y} - \sqrt{ y_k } \right) + \mu_o , 
\label{pixamp}
\end{equation}
where $\Delta y$ is the length of the side perpendicular to the fold
caustic. Throughout $\mu_o = 1$.

The ``strength'' of the caustic, $g$, was determined from a numerical
study undertaken by Witt (1990). In this study, the parameter $\hat{K} =
\sqrt{g}$, and, choosing a medium microlensing optical depth $(\sigma\sim
0.2)$, $\hat{K}$ was chosen to be 0.1, 0.4, 0.7, and 1.0. These values are
representative of the broad distribution in caustic strengths found
at such median optical depths.

For these various ``strengths'', the caustic was swept across the
source and the resultant flux, as a function of wavelength, was
calculated.  The resulting spectrum was normalized to the flux of the
amplified unattenuated continuum.  With this, the microlensed and
intrinsic form of the BAL profile could be compared.

\section{Results}\label{results}

\begin{table*}
\begin{center}
\begin{tabular}{crrrrrrrrr}
\hline\hline
Cloud Size & \multicolumn{4}{c}{$\hat{K}=0.1$} & & \multicolumn{4}{c}{$\hat{K}=0.4$} \\
\cline{2-5}\cline{7-10}
$2^{-n}$ & 5\% &15\% &25\% &50\% & &5\% &15\% &25\% & 50\%  \\
\hline
1   & 2.14 & 5.56 & 6.46 & 11.02 & & 4.65 & 12.07 & 14.06 & 23.95 \\ 
2   & 1.17 & 3.11 & 7.39 &  5.37 & & 2.54 &  8.63 & 16.07 & 11.81 \\ 
3   & 0.93 & 2.35 & 3.88 &  4.18 & & 2.23 &  5.63 &  9.33 & 10.03 \\
4   & 0.55 & 1.32 & 2.05 &  2.12 & & 1.20 &  3.10 &  4.64 &  5.11 \\ 
5   & 0.29 & 0.82 & 1.15 &  1.72 & & 0.65 &  1.87 &  2.84 &  3.91 \\ 
6   & 0.16 & 0.49 & 0.74 &  0.76 & & 0.38 &  1.14 &  1.68 &  1.90 \\ 
7   & 0.09 & 0.24 & 0.36 &  0.51 & & 0.23 &  0.61 &  0.87 &  1.06 \\ 
8   & 0.05 & 0.16 & 0.25 &  0.23 & & 0.12 &  0.38 &  0.53 &  0.58 \\ 
9   & 0.03 &  --- & 0.11 &  0.15 & & 0.08 &  ---  &  0.26 &  0.34 \\ 
\hline\hline
Cloud Size & \multicolumn{4}{c}{$\hat{K}=0.7$} & & \multicolumn{4}{c}{$\hat{K}=1.0$} \\
\cline{2-5}\cline{7-10}
$2^{-n}$ & 5\%&15\%&25\%&50\%& &5\%&15\%&25\% & 50\% \\
\hline
1   & 5.59 & 14.50 & 16.91 & 28.77 & &6.08 & 15.76 & 18.40 & 31.29 \\
2   & 3.33 & 11.57 & 19.31 & 15.84 & &3.85 & 13.40 & 21.56 & 18.34 \\
3   & 2.78 &  7.20 & 11.67 & 12.55 & &3.09 &  8.34 & 12.96 & 14.29 \\
4   & 1.44 &  4.25 &  5.82 &  6.49 & &1.67 &  5.08 &  6.47 &  7.81 \\
5   & 0.79 &  2.31 &  3.60 &  4.78 & &1.03 &  2.60 &  4.04 &  5.25 \\
6   & 0.50 &  1.42 &  2.11 &  2.48 & &0.58 &  1.70 &  2.39 &  2.81 \\
7   & 0.30 &  0.81 &  1.09 &  1.25 & &0.33 &  0.92 &  1.24 &  1.37 \\
8   & 0.16 &  0.48 &  0.63 &  0.74 & &0.18 &  0.54 &  0.68 &  0.84 \\
9   & 0.10 &   --- &  0.34 &  0.44 & &0.11 &   --- &  0.39 &  0.50 \\
\hline\hline 
\end{tabular}
\caption[]{ Maximum deviations in the lensed absorption spectra for four 
caustic strengths, $\hat{K}=0.1$, 0.4, 0.7, and 1.0, and four
absorption dispersions, $\omega=5\%$, $15\%$, $25\%$, and $50\%$.  The
cloud size is represented as $2^{-n}r_{s}, n=1..9$, where $r_{s}=
3.7$ source radii, or $3.7\times10^{15}{\rm h_{75}^{-1}cm}$, and ranges
from $7.1\times10^{12}{\rm h_{75}^{-1}cm} - 1.8\times10^{15}{\rm
h_{75}^{-1}cm}$ in extent. The deviations are given as percentages of
the normalized continuum flux, while the absorption dispersions are
percentages of total absorption (see Section~\ref{model}). }
\label{table_maxdev}
\end{center}
\end{table*}

\begin{table}
\begin{center}
\begin{tabular}{crrrrr}
\hline\hline
Cloud Size   &  \multicolumn{5}{c}{Absorption Dispersion ($\omega$)}\\
\cline{2-6}
$2^{-n}$ & 15\%	& 20\%    & 25\%    & 30\%    & 40\%  \\
\hline
1       & 15.76	& 16.75   & 18.40   & 18.43   & 27.48  \\
2       & 13.40	& 17.60   & 21.56   & 19.01   & 24.94  \\
3       &  8.34	& 10.64   & 12.96   & 12.57   & 15.97  \\
4       &  5.08	&  6.45   &  6.47   &  6.64   &  8.62 \\
5       &  2.60	&  3.70   &  4.04   &  4.55   &   --- \\
\hline\hline
\end{tabular}
\caption[]{Maximum deviations for caustic strength $\hat{K}=1.0$ and 
varying absorption dispersions ($\omega$).  Cloud size is given as
$2^{-n}r_{s}, n=1..5$ (see Table~\ref{table_maxdev}), deviations
are given as a percentage of the normalized continuum flux, and the
absorption dispersion is the percentage of total absorption (see
Section~\ref{model}).}
\label{table_maxdevk1}
\end{center}
\end{table}

As the caustic line sweeps across the source, the magnification of the
flux as seen through each ``cloud'' is calculated using
Equation~\ref{pixamp} and summed to give the total flux as a function
of wavelength.  As the entire flux from the region is amplified, the
microlensed spectra are renormalized such that the continuum flux is
one.  The results of this analysis are summarized in
Table~\ref{table_maxdev}, which presents the maximum flux deviations
(as a percentage of the normalized continuum flux) between the
unlensed spectra and lensed spectra. For large clouds, significant
deviations from the unlensed form of the BAL profile can be seen, with
these differences increasing with the caustic strength, $\hat{K}$, and
the dispersion in the absorption, $\omega$.

Importantly, the degree of deviation is strongly dependent on the
cloud size, with small clouds producing the smallest deviations, even
with significantly inhomogeneous absorption distributions $(\omega =
50\%)$.  In this regime even the high amplification of an individual
cloud is washed out in the overall amplification of the region. For
such cloud distributions the BAL absorption region is essentially seen
as smooth.

The solid line in Figure~(3) presents an example of a microlensed BAL
spectrum, while the dashed line represents the unlensed form of the
BAL profile.  As can be seen, there is significant deviations of the
microlensed spectrum from the unlensed source ($\sim30\%$).  This is
of similar order to the deviations observed in the spectra of
H1413+117 [see Figure (1) in Hutsem\'ekers (1993)].  In this case,
the absorption clouds are $1.8 \times 10^{15}{\rm h_{75}^{-1}cm}$ in
extent ($2^{-1}r_{s}$), with an absorption dispersion $\omega=50\%$,
while the lensing caustic is given a strength of $\hat{K}=1.0$.

\begin{figure}
\centerline{
\psfig{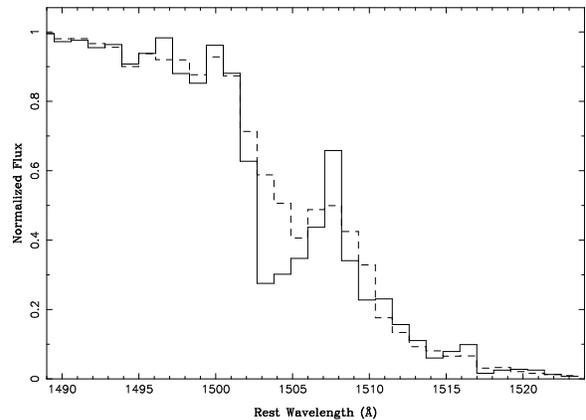}
}
\label{spectrum}
\caption[]{ 
The lensed absorption spectrum of a source with a cloud size  
of $1.8\times10^{15}{\rm h_{75}^{-1}cm}$ ($2^{-1}r_{s}$),
an absorption dispersion $\omega= 50\%$, and a lensing caustic strength
$\hat{K}=1.0$.  The solid line represents the lensed spectrum at 
its maximum deviation of 31\% (see Table~\ref{table_maxdev}) from the 
unlensed spectrum (dashed line).  Each pixel has a velocity resolution
of $\sim200km\;s^{-1}$.  }
\end{figure}

Figure~(4) presents a typical amplification curve, demonstrating the
temporal nature of variations. The cloud size in this simulation is
$5.7\times10^{13}{\rm h_{75}^{-1}cm}$ ($2^{-6}r_{s}$), with an
absorption dispersion of $\omega=50\%$, while the caustic had a
strength $\hat{K}=1.0$.  The solid line presents the maximum
deviation, at any wavelength, between the normalized microlensed
spectrum and the unlensed profile, expressed as the absolute deviation
from the continuum flux. The dashed line similarly presents the
average absolute deviation as the caustic sweeps across the source.
The degree of the deviation is strongly dependent on the the total
amplification of the source, being zero before the caustic crosses,
and tending again to zero after the caustic has passed.

\begin{figure}
\centerline{
\psfig{figure=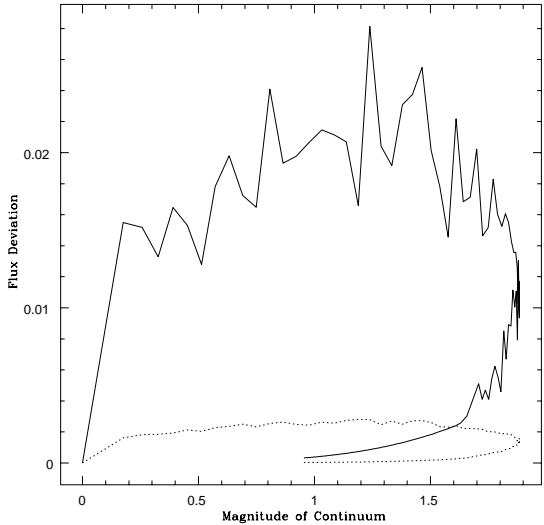,width=3.0in,angle=0}
}
\label{curve}
\caption[]{ 
A typical magnification curve for a lensed source with absorption
clouds of $5.7\times10^{13}{\rm h_{75}^{-1}cm}$ ($2^{-6}r_{s}$),
$\omega=50\%$, and $\hat{K}=1.0$.  The solid line represents the
maximum deviation of lensed from unlensed source at each step of the
caustic across the source.  The dashed line is the average deviation
as the caustic sweeps across the source.  As the passage of the
caustic over the source is reversible, these loops can be followed
around either way.}
\end{figure}

As the cloud size approaches and becomes larger than the continuum
region itself, the distribution of the absorbing material in front of
the quasar becomes homogeneous.  Such a scenario would not result in
any flux deviations as the caustic sweeps over the source.  We
therefore would expect a turnover in the degree of fluctuations as the
cloud size approaches this limit.  To test this, additional
simulations were undertaken for a fixed caustic strength,
$\hat{K}=1.0$, varying absorption dispersions, $\omega=15\%$, $20\%$,
$25\%$, $30\%$, and $40\%$, and cloud sizes of $2^{-n}r_{s}, n=1..5$.
Results of this are summarized in Table~\ref{table_maxdevk1}.  This
turnover is not apparent for small absorption dispersions, but is seen
for medium absorption values.  The turnover again vanishes for large
absorption dispersions.  This could be a result of the variation of
$\omega$ (see Section~\ref{model}) at the extreme ends of the BAL
profile.  For these absorption dispersions, $\omega$ is constant for
the range $\sim2\Delta\lambda$, whereas for the larger or smaller
absorption dispersions, $\omega$ is constant for a range that is much
greater or less than $2\Delta\lambda$.  This affect could also be due to a
combination of both cloud size and absorption dispersion, as we see
that the deviation for a $2^{-2}r_{s}$ cloud size and $\omega=25\%$ is
greater than that for the same cloud size and a larger absorption
dispersion, $\omega=30\%$.

From this analysis of the microlensing of BAL quasars by single caustic
crossings, it is seen that the largest deviations between the lensed
and unlensed spectra are a result of a large cloud size and a large
absorption dispersion. The maximum deviations do not depend on the
caustic strength as strongly.

\section{Comparison to Previous Models}

The models of Hutsem\'ekers (1993, Hutsem\'ekers et al. 1994)
considered the action of an individual microlens on a single cloud, or
inhomogeneity, within the BAL region. Here, the single cloud
approximation has been discarded and the entire BAL region consists of
an ensemble of such inhomogeneities in front of the continuum
source. With this, the need for a specific cloud to be placed in a
``fine--tuned'' position with respect to the continuum is no longer
necessary; each cloud in the BAL region is amplified individually as
the caustic sweeps across it, and the total flux from the region is
the sum of these sub--regions. With disparate absorption properties,
characterised by $\omega$, the stocastic positioning of the clouds
ensures that, for larger cloud scale--sizes, significant spectral
variability will result during a caustic crossing.

Furthermore, the analysis presented here considered caustic strengths
drawn from an ensemble of microlensing masses at moderate optical
depths.  This situation more accurately represents the situation in
the macrolensed BAL quasar, H1413+117, where the optical depths to the
images are expected to be significant~\cite{kayser1990} and the
isolated lens approximation is no longer applicable.

\section{Conclusions}\label{conclusions}

This paper has presented an analysis of the microlensing effects of a
single line caustic as it sweeps across a broad absorption line region
in front of a quasar continuum source.  Our results demonstrate that
significant spectral variations can be induced during such an event,
unlike previous results which required ''fine-tuned'' models.  The
identification of these variations provides an effective tool in
determining the scale of structure of BAL regions of quasars, because
the physical nature, both size and amount of absorption, of the clouds
is directly related to the variations in the flux of the lensed
object. 

Further monitoring of H1413+117 is important.  Our numerical results
show microlensing can produce spectroscopic differences of up to
$\sim~30\%$ which would be observable within the spectra of the
multiple images of H1413+117.  Monitoring over time scales of several
months will allow a separation of intrinsic variability, which will
be correlated amongst the images, and microlensing induced
variability, which, although occurring on similar time scales, will be
apparent in only individual images.  Also, with microlensing,
spectroscopic deviations will be strongly correlated with the apparent
image flux.

The analysis presented here suggests that, if the spectral deviations
observed in this system are due to the action of gravitational
microlensing, then the cloud size must be a substantial fraction of
the size of the continuum source.  With our adopted source profile
this implies that the clouds are of order $10^{14}-10^{15}$cm in
extent which is an order of magnitude larger than suggested in
previous studies (see Section 2.2).  One possible solution to this
inconsistency is that clouds actually do possess a small physical size
but are clumped on larger scales. 

A more comprehensive study, considering differing kinematic and
spatial structures in the BAL region, as well as the effects of an
ensemble of microlenses of various optical depths, velocity
dispersions, and mass functions, is the subject of a forthcoming paper
(Belle and Lewis, in preparation).

\section{Acknowledgments}
We thank Paul Hewett and Craig Foltz for engaging discussions on the
nature of BAL QSOs.  We are also grateful to the referee, Dr Prasenjit
Saha, for very useful comments.

\end{document}